\documentclass[10pt,conference]{IEEEtran}

\usepackage{graphicx} 
\usepackage{color}
\usepackage{psfrag}
\usepackage{epsfig}

\definecolor{Red}{rgb}{1,0,0}
\definecolor{Blue}{rgb}{0,0,1}
\definecolor{Olive}{rgb}{0.41,0.55,0.13}
\definecolor{Green}{rgb}{0,1,0}
\definecolor{MGreen}{rgb}{0,0.8,0}
\definecolor{DGreen}{rgb}{0,0.55,0}
\definecolor{Yellow}{rgb}{1,1,0}
\definecolor{Cyan}{rgb}{0,1,1}
\definecolor{Magenta}{rgb}{1,0,1}
\definecolor{Orange}{rgb}{1,.5,0}
\definecolor{Violet}{rgb}{.5,0,.5}
\definecolor{Purple}{rgb}{.75,0,.25}
\definecolor{Brown}{rgb}{.75,.5,.25}
\definecolor{Grey}{rgb}{.5,.5,.5}

\usepackage{booktabs} 
\usepackage{array} 
\usepackage{paralist} 
\usepackage{verbatim} 
\usepackage{subfigure} 
\usepackage{amsmath,amssymb}

\newtheorem{theorem}{Theorem}

\newtheorem{lemma}{Lemma}

\newtheorem{example}{Example}

\newtheorem{remark}{Remark}

\newcommand{\p}{{\rm P}}

\def\cX{{\cal X}}
\def\cY{{\cal Y}}

\begin{document}

\title{Comments on ``Broadcast Channels with Arbitrarily Correlated Sources''}

\author{
\authorblockN{Gerhard Kramer}
\authorblockA{Department of Electrical Engineering\\
University of Southern California \\
Los Angeles, CA, USA\\
Email: gkramer@usc.edu}
\and
\authorblockN{Chandra Nair}
\authorblockA{Department of Information Engineering\\
Chinese University of Hong Kong \\
Sha Tin, N.T., Hong Kong\\
Email: chandra@ie.cuhk.edu.hk}
}

\maketitle



\begin{abstract}
The Marton-Gelfand-Pinsker inner bound on the capacity region of broadcast channels
was extended by Han-Costa to include arbitrarily correlated sources where the capacity
region is replaced by an admissible source region. The main
arguments of Han-Costa are correct but unfortunately the authors overlooked
an inequality in their derivation. The corrected region is presented and the absence
of the omitted inequality is shown to sometimes admit sources that are not admissible.
\end{abstract}


\section{Introduction}
We borrow terminology from~\cite{HanCosta87} with minor modifications. Consider a
two-receiver broadcast channel (BC), say $\omega$, with correlated, or more precisely
dependent, sources $(S,T)$. Source $(S,T)$ is said to be {\it admissible} for this BC if for any
$\lambda$, $0<\lambda<1$, and for large enough $n$ there is a code with length-$n$
codewords such that $Pe_1\le\lambda$ and $Pe_2\le\lambda$, where $Pe_1$ and $Pe_2$ are
the respective error probabilities for receivers 1 and 2. The set of all admissible sources
is called the {\it admissible source region}.

Han and Costa developed a (purported) subset of the admissible source region for BCs
with arbitrarily correlated sources in~\cite[Theorem 1 and Example 1]{HanCosta87}.
We observe that the main arguments in \cite{HanCosta87} are valid but the authors unfortunately
overlooked an inequality in one of the final steps of their proof.
The corrected versions of~\cite[Theorem 1 and Example 1]{HanCosta87} are presented in
Sec.~\ref{sec:correction}.
On the other hand, since the admissible source region is not known in general, it is not a priori
clear whether or not the Han-Costa source set is in fact a subset of the admissible
source region after all. We rule out this possibility by giving two examples where the
Han-Costa source set includes sources that are not admissible.

\section{Revised Theorem 1 and Example 1 in \cite{HanCosta87}}
\label{sec:correction}

The wording of the theorem and the example below are taken with minor
modifications from \cite{HanCosta87}.

\begin{theorem}[revised from~\cite{HanCosta87}]
\label{th:main}
Suppose that a broadcast channel $\omega$ and a source $(S,T)$ are given, and let
$K = f(S) = g(T)$ be the common variable in the sense of Gacs and K\"{o}rner (and also
Witsenhausen). If there exist auxiliary random variables $W,U,V$ (with values in finite sets)
that satisfy the Markov chain property
\begin{align}
  ST - WUV - X - Y_1 Y_2  \label{eq:thm�1-chain}
\end{align}
and the inequalities
\begin{align} 
H(S) &\leq I(SWU;Y_1) - I(T;WU|S) \label{eq:thm1-1} \\
H(T) &\leq I(TWV;Y_2) - I(S;WV|T)  \label{eq:thm1-2}  \\
H(ST) & \leq \min\{I(KW;Y_1), I(KW;Y_2)\} + I(SU;Y_1|KW) \nonumber \\
&\quad  + I(TV;Y_2|KW) - I(SU;TV|KW) \label{eq:thm1-3} \\
H(ST) & \leq I(SWU;Y_1) + I(TWV;Y_2) - I(SU;TV|KW) \nonumber \\
&\quad - I(ST;KW). \label{eq:thm1-4}
\end{align}
then the source $(S,T)$ is admissible for the channel $\omega$. Here, $X$ is an input variable
with values in the input alphabet $\cX$, and $Y_1,Y_2$ are the output variables with values
in the output alphabets $\cY_1,\cY_2$, respectively, induced by $X$ via $\omega$.
\end{theorem}

\begin{example}[revised from~\cite{HanCosta87}]
\label{ex:1}
Consider sources with $S=(S_0,K), T=(T_0,K)$, where $S_0,T_0,K$
are statistically independent, and where $H(K)=R_0$, $H(S_0)=R_1$, $H(T_0)=R_2$.
If we choose $WUV$ to be independent of $ST$, then the conditions of
Theorem \ref{th:main} reduce to the Markov chain property
\begin{align}
  WUV - X - Y_1 Y_2 \label{eq:ex1-chain}
\end{align}
and the inequalities
\begin{align}
R_0 + R_1 &< I(WU;Y_1) \label{eq:ex1-1} \\
R_0 + R_2 &< I(WV;Y_2) \label{eq:ex1-2} \\
R_0 + R_1 + R_2 &< \min\{I(W;Y_1), I(W;Y_2)\} + I(U;Y_1|W) \nonumber \\
&\quad + I(V;Y_2|W) - I(U;V|W) \label{eq:ex1-3}\\
2R_0 + R_1 + R_2 &< I(WU;Y_1) + I(WV;Y_2) - I(U;V|W). \label{eq:ex1-4}
\end{align}
\end{example}

\begin{remark}
Inequalities (\ref{eq:thm1-4}) and (\ref{eq:ex1-4}) are missing in \cite{HanCosta87}.
Note that the revised Example 1 is a special case of a more general result
that appeared in the Ph.D.\ thesis of Y.\ Liang in 2005
(see \cite[p.~89, Remark 10]{LiangThesis} and \cite[Theorem 5]{Liang07}).
Note also that the rate region (\ref{eq:ex1-chain})-(\ref{eq:ex1-4}) was shown to be
equivalent to the Marton-Gelfand-Pinsker region in \cite{LiangAllerton08}
(what we call the ``Marton-Gelfand-Pinsker region'' is given in~\cite[Theorem~1]{Gelfand80}
and~\cite[p.~391, Problem~10(c)]{Csiszar81}).
\end{remark}

\subsection{Solving the Case of the Missing Inequality}

In \cite[p.~647]{HanCosta87}, the authors derive the following valid inequalities,
see Equations (3.34)-(3.37):
\begin{align*}
 & H(S|KW) + H(T|KW) + H(K) \\
&\qquad < I(TVW;Y_2) + I(SU;Y_1|KW) - \rho_0 -\rho_1 - \rho_2 \\ 
&H(T|KW) + H(K) < I(TVW;Y_2)  - \rho_0  - \rho_2 \\
 &H(S|KW) + H(T|KW) + H(K) \\
&\qquad < I(SUW;Y_1) + I(TV;Y_2|KW) - \rho_0 -\rho_1 - \rho_2 \\ 
&H(S|KW) + H(K) < I(SUW;Y_1)  - \rho_0  - \rho_1.
\end{align*}
They next eliminate the variables $\rho_0, \rho_1, \rho_2$ using the following inequalities
in~\cite[p.~645]{HanCosta87}, see Equations (3.5)-(3.8):
\begin{align*}
 \rho_0 &> I(ST;W|K) \\
\rho_1 &> I(T;U|SW) \\
\rho_2 &> I(S;V|TW) \\
\rho_1 + \rho_2 & > I(SU;TV|W) - I(S;T|W). 
\end{align*}
The oversight occurs in this elimination. 
Eliminating $\rho_0,\rho_1, \rho_2$ and removing redundant inequalities
we obtain the bounds in~\cite[(3.38)-(3.41)]{HanCosta87}:
\begin{align}
 & H(S|KW) + H(T|KW) + H(K) + I(ST;W|K) \nonumber \\
& \qquad < I(TVW;Y_2) + I(SU;Y_1|KW)  - A  \\
& H(T|KW) + H(K) \nonumber \\
& \qquad < I(TVW;Y_2) - I(ST;W|K) - I(S;V|TW) \medskip \\
& H(T|KW) + H(S|KW) + H(K) + I(ST;W|K) \nonumber \\
& \qquad < I(SUW;Y_1) + I(TV;Y_2|KW)  - A \\
& H(S|KW) + H(K) \nonumber \\
& \qquad < I(SUW;Y_1) - I(ST;W|K) - I(T;U|SW)
\end{align}
where $A=I(SU;TV|W)-I(S;T|W)$, as well as the bound
\begin{align}
&H(S|KW) + H(T|KW) + 2H(K) + 2I(ST;W|K) \nonumber \\
& \qquad < I(SUW;Y_1) + I(TVW;Y_2) - A. \label{eq:missing}
\end{align}
It is the inequality (\ref{eq:missing}) that was omitted in~\cite{HanCosta87}.

Continuing as in~\cite[p.~647]{HanCosta87}, we obtain the revised Theorem~1
by using the equalities
\begin{align*}
& H(S|KW) + H(K) = H(S) - I(S;W|K)\\
& H(T|KW) + H(K) = H(T) - I(T;W|K) \\
& H(S|KW) + H(T|KW) + H(K) \nonumber \\
& \quad =  H(ST) + I(S;T|K)- I(S;W|K) - I(T;W|K).
\end{align*}

\subsection{Counterexample~\ref{sse:exkramer}}
\label{sse:exkramer}

Since the set-up of Example 1 is a well-studied and important case, we explore the
following question: If we remove the inequality (\ref{eq:ex1-4}) then is the resulting rate
region (the Han-Costa region of ~\cite[Example~1]{HanCosta87}) always achievable?
We develop two counterexamples to show that this is not the case. The reader will notice
that the counterexamples are closely related. We present them both for reasons 
that will become clear in Sec.~\ref{sec:historical}.

As a first counterexample, consider the deterministic BC
\begin{align}
  (Y_1,Y_2) = \left\{ \begin{array}{ll}
     (0,0), & X=0 \\
     (1,0), & X=1 \\
     (1,1), & X=2 \\
     (2,1), & X=3 . \end{array} \right.
     \label{eq:channel1}
\end{align}
The capacity region of a deterministic BC is known to be
the union over Markov chains $W-X-Y_1Y_2$ of the non-negative
rate triples $(R_0,R_1,R_2)$ satisfying (see~\cite[p.~391]{Csiszar81})
\begin{align}
   & R_0 \le \min\left( I(W;Y_1), I(W;Y_2) \right) \label{eq:R0-bound} \\
   & R_0+R_1 \le H(Y_1) \\
   & R_0+R_2 \le H(Y_2) \\
   & R_0+R_1+R_2 \le \min \left( I(W;Y_1), I(W;Y_2) \right) + H(Y_1 Y_2|W). 
      \label{eq:sum-rate-bound}
\end{align}
We mimic the development of~\cite[Sec.~IV]{Ratnakar:06}. Suppose we would like to achieve
\begin{align}
  R_0+R_1+R_2 = H(Y_1Y_2) \label{eq:sum-rate}
\end{align}
for the BC (\ref{eq:channel1}). For example, we can achieve $(R_0,R_1,R_2)=(0,1,1)$
and $H(Y_1 Y_2)=2$ by choosing $W$ to be a constant and $X$ uniform. It is easy to check that for
(\ref{eq:sum-rate}) to be satisfied, one must have the double Markov relations
\begin{align}
   & W-Y_1-Y_2 \\
   & W-Y_2-Y_1
\end{align}
in the expression (\ref{eq:sum-rate-bound}).

Suppose next that we would like to achieve
\begin{align}
  R_0+R_1+R_2 = 2
\end{align}
for the BC (\ref{eq:channel1}), as in the example we just considered.
Obviously, the input $X$ must be uniform, and for this choice of
$X$ one can check that the joint distribution of $(Y_1,Y_2)$ is indecomposable in
the sense of \cite[p.~350]{Csiszar81}. This further implies, by  \cite[p.~402]{Csiszar81}, that
\begin{align}
  W \text{ is independent of } Y_1Y_2 \label{eq:independence}
\end{align}
and therefore, by (\ref{eq:R0-bound}), that $R_0=0$. One can further check that with uniform $X$,
but without the bound (\ref{eq:R0-bound}), the following rate-triple is permitted
\begin{align}
  (R_0,R_1,R_2)=(1/2,1,1/2) \label{eq:rate-triple}
\end{align}
Thus, the bound (\ref{eq:R0-bound}) is needed because the rate-triple
(\ref{eq:rate-triple}) is not achievable.

Finally, note that we are further suggesting that one replace (\ref{eq:R0-bound}) with the
bound (\ref{eq:ex1-4}) where $U=Y_1$ and $V=Y_2$, i.e, with
\begin{align}
   2R_0+R_1+R_2 \le I(W;Y_1) + I(W;Y_2) + H(Y_1Y_2|W).  \label{eq:2R0-bound}
\end{align}
For example, if $R_0+R_1+R_2=2$ then from (\ref{eq:independence}) 
and (\ref{eq:2R0-bound}) we see that we must have
\begin{align}
   2R_0+R_1+R_2 \le H(Y_1 Y_2) = 2
\end{align}
so that $R_0=0$. Summarizing, we need to add the bound (\ref{eq:R0-bound})  {\it or}
the bound (\ref{eq:ex1-4}) to the bounds (2.14)-(2.16) in~\cite{HanCosta87}.
The equivalence of adding either bound was proved for general broadcast
channels in~\cite{LiangAllerton08}.

\subsection{Counterexample~\ref{sse:exnair}}
\label{sse:exnair}

\begin{figure}[ht]
\begin{center}
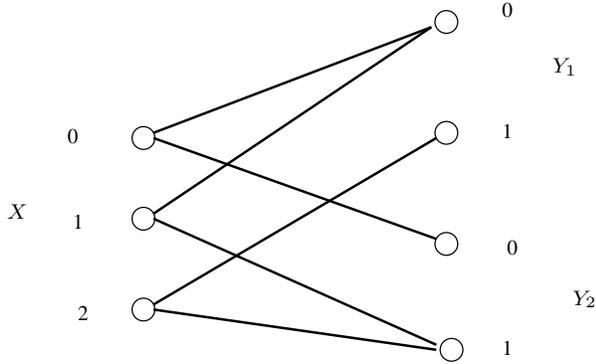
\caption{Blackwell Channel}
\label{fig:blackwell}
\end{center}
\end{figure}

Consider the Blackwell BC shown in Fig.~\ref{fig:blackwell}.
This channel is deterministic so the capacity region is given by
(\ref{eq:R0-bound})-(\ref{eq:sum-rate-bound}) where $W-X-Y_1Y_2$ forms a
Markov chain. We have the following lemma that is closely related to
Counterexample~\ref{sse:exkramer}.

\begin{lemma}
\label{lem:main}
If $W - X - Y_1Y_2$ forms a Markov chain and 
\begin{equation*}
\min\left( I(W;Y_1),I(W;Y_2) \right) + H(Y_1Y_2|W) = H(Y_1Y_2)
\end{equation*}
for the Blackwell channel in Fig.~\ref{fig:blackwell}  then the random variables $W$ and $X$
are independent.
\end{lemma}

Before we prove Lemma \ref{lem:main}, we claim that the lemma provides a counterexample
to our question posed above. In particular, the lemma implies that if
$R_0 + R_1 + R_2 \to H(Y_1Y_2)$, then we must have
$R_0 \big( \leq \min\left( I(W;Y_1), I(W;Y_2) \right) \big) \to 0$. Thus the triple
$R_0 = I(Y_1;Y_2) - \epsilon$,
$R_1 = H(Y_1|Y_2), R_2 = H(Y_2|Y_1)$, is not achievable for $\epsilon$ small enough. However,
this rate-triple is permitted by~\cite[Example 1]{HanCosta87}.

\begin{proof}(Lemma \ref{lem:main})
The equality
$$ \min\left( I(W;Y_1),I(W;Y_2) \right) + H(Y_1Y_2|W) = H(Y_1Y_2) $$ 
implies that $I(W;Y_1) = I(W;Y_2) = I(W;Y_1Y_2)$. Using the observation that for a Blackwell
channel $X$ is a deterministic function of $Y_1Y_2$, we have the following equalities
\begin{align}
I(W;Y_1) &= I(W;X) \label{eq:con1} \\
I(W;Y_2) &= I(W;X) \label{eq:con2}.
\end{align}

From \eqref{eq:con1} and the Markov relationship $W - X - Y_1$ we see that
$I(W;X|Y_1) = 0$ and therefore
\begin{align}
  \p(X=0|Y_1=0,W=w) = \p(X=0|Y_1=0). \label{eq:xyw}
\end{align}
For the Blackwell Channel, \eqref{eq:xyw} is equivalent to
\begin{align*}
   &\frac{\p(X=0)}{\p(X=0) + \p(X=1)} \\
   &\quad = \frac{\p(X=0|W=w)}{\p(X=0|W=w) + \p(X=1|W=w)}.
\end{align*} 
Thus we obtain
\begin{equation}
\label{eq:con3}
\frac{\p(X=0)}{\p(X=1)} = \frac{\p(X=0|W=w)}{\p(X=1|W=w)}.
\end{equation}

Similarly starting from \eqref{eq:con2} and the Markov relationship
$W - X - Y_2$, we compute
\begin{equation}
\label{eq:con4}
\frac{\p(X=2)}{\p(X=1)} = \frac{\p(X=2|W=w)}{\p(X=1|W=w)}.
\end{equation}
From \eqref{eq:con3} and \eqref{eq:con4} we deduce
\begin{equation*}
 \p(X=i|W=w) = \p(X=i), ~\mbox{for}~i=0,1,2.
\end{equation*}
 This concludes the proof of Lemma \ref{lem:main}. 
\end{proof}

\section{Historical Remarks}
\label{sec:historical}

The revised Theorem 1 and Example~\ref{sse:exkramer} were developed by G.\ Kramer
in the summer of 2005. His motivation was that S.\ Shamai pointed out to him that the
potential improvement (\ref{eq:ex1-chain})-(\ref{eq:ex1-4})
of the Marton-Gelfand-Pinsker region that appeared in the Ph.D.\ thesis of Y.\ Liang
(see \cite[p.~89, Remark 10]{LiangThesis} and \cite[Theorem 5]{Liang07}) was superseded
by the earlier results of Han-Costa~\cite{HanCosta87}. 
Kramer communicated the revised Theorem 1 and Example~\ref{sse:exkramer}  to Shamai
and Han in August 2005 via email but did not otherwise document the results.

In 2008, Y.-H.\ Kim queried C.\ Nair about the validity of the results in Han-Costa~\cite{HanCosta87}.
Nair independently discovered and corrected the error of~\cite{HanCosta87} in 2008 and
developed Example~\ref{sse:exnair}. He forwarded a write-up of his results to A.\ El Gamal and
M.\ Costa. Costa forwarded the write-up to Han, who then replied back with the earlier
communication by Kramer. This eventually led to the current joint paper.

\section*{Acknowledgement}
The authors wish to thank S.\ Shamai, T.\ S.\ Han, M.\ Costa, A.\ El Gamal, and
Y.-H.\ Kim for their roles in bringing these results together.

\bibliographystyle{IEEEtran}
\bibliography{cap}

\begin{thebibliography}{1}
\providecommand{\url}[1]{#1}
\csname url@samestyle\endcsname
\providecommand{\newblock}{\relax}
\providecommand{\bibinfo}[2]{#2}
\providecommand{\BIBentrySTDinterwordspacing}{\spaceskip=0pt\relax}
\providecommand{\BIBentryALTinterwordstretchfactor}{4}
\providecommand{\BIBentryALTinterwordspacing}{\spaceskip=\fontdimen2\font plus
\BIBentryALTinterwordstretchfactor\fontdimen3\font minus
  \fontdimen4\font\relax}
\providecommand{\BIBforeignlanguage}[2]{{%
\expandafter\ifx\csname l@#1\endcsname\relax
\typeout{** WARNING: IEEEtran.bst: No hyphenation pattern has been}%
\typeout{** loaded for the language `#1'. Using the pattern for}%
\typeout{** the default language instead.}%
\else
\language=\csname l@#1\endcsname
\fi
#2}}
\providecommand{\BIBdecl}{\relax}
\BIBdecl

\bibitem{HanCosta87}
T.~S. Han and M.~H.~M. Costa, ``Broadcast channels with arbitrarily correlated
  sources,'' \emph{IEEE Trans.\ Inf.\ Theory}, vol.~33, no.~5, pp. 641--650,
  Sep. 1987.

\bibitem{LiangThesis}
Y.~Liang, ``Multiuser communications with relaying and user cooperation,''
  \emph{\textup{Ph.D. dissertation, Department of Electrical and Computer
  Engineering, University of Illinois at Urbana-Champaign, Illinois}}, 2005.

\bibitem{Liang07}
Y.~Liang and G.~Kramer, ``Rate regions for relay broadcast channels,''
  \emph{IEEE Trans.\ Inf.\ Theory}, vol.~53, no.~10, pp. 3517--3535, Oct. 2007.

\bibitem{LiangAllerton08}
Y.~Liang, G.~Kramer, and H.~V. Poor, ``Equivalence of two inner bounds on the
  capacity region of the broadcast channel,'' in \emph{Proc. 46th Annual
  Allerton Conf. on Commun., Control, and Computing}, Monticello, IL, Sept.
  23--26 2008.

\bibitem{Gelfand80}
S.~I. Gel'fand and M.~S. Pinsker, ``Capacity of a broadcast channel with one
  deterministic component,'' \emph{Probl. Inform. Transm.}, vol.~16, no.~1, pp.
  17--25, Jan.-Mar. 1980.

\bibitem{Csiszar81}
{I. Csisz\'ar and J. K\"orner}, \emph{Information Theory: Coding Theorems for
  Discrete Memoryless Channels}.\hskip 1em plus 0.5em minus 0.4em\relax
  Budapest: Akad\'emiai Kiad\'o, 1981.

\bibitem{Ratnakar:06}
N.~Ratnakar and G.~Kramer, ``The multicast capacity of deterministic relay
  networks with no interference,'' \emph{IEEE Trans.\ Inf.\ Theory}, vol.~52,
  no.~6, pp. 2425--2432, June 2006.

\end{thebibliography}

\end{document}